\documentstyle[aps,epsf]{revtex}
\begin{document}
\draft
\title{A theory of electromagnetism with uniquely defined potential
and covariant conserved spin.}
\vskip 0.5 truecm 
\author {A. B. van Oosten\cite{email}}
\vskip 0.3cm
\address{Laboratoire de Physique Quantique,
IRSAMC, Universit\'e Paul Sabatier  
31062 Toulouse, France\\ }
\date{Preprint \today}
\maketitle

\vskip0.3cm

\begin{abstract}

The Lagrangian 
$ {1 \over 2} \epsilon_0 c^2 \partial_\mu A_\nu \partial^\mu A^\nu $
is shown to yield a non-gauge-invariant theory of electromagnetism.
The potential is uniquely determined
by the inhomogeneous wave equation
and boundary conditions at infinity.
The Lorenz condition and minimal coupling follow from charge conservation.
Electromagnetic spin is conserved and a spin operator is proposed
without sacrificing covariance.
Covariant quantisation is carried out without redefining the metric.
It is a valid alternative to the standard approach
since it makes the same experimental predictions. 

\end{abstract}

\vskip 0.3cm
\vskip 0.3cm

Maxwell's equations express electromagnetism in terms of the electric
and magnetic fields, $F_{\mu\nu}$. They imply that a potential field $A_\mu$ can
be defined through $F_{\mu\nu} = \partial_\mu A_\nu - \partial_\nu A_\mu$, but
only up to a contribution of the form $\partial_\mu \chi$, with $\chi$ an undefined
scalar field. In spite of its incomplete definition, it is $A_\mu$ and its
derivatives, not $F_{\mu\nu}$, that constitute the functional variables in
Lagrangian formalism. The equations can be derived
from the standard\cite{gmunu} gauge-invariant
\begin{equation}
{\cal L}_{ST} = {1 \over 4} \epsilon_0 c^2 F_{\mu\nu} F^{\mu\nu} 
\label{lmax}
\end{equation}
or any equivalent Lagrangian differing from it by a four-divergence.
Such Lagrangians, however, yield non-gauge-symmetric Noether
conservation laws, asymmetric energy-momentum tensors, non-conserved spin and
disallow quantisation. Interaction with matter is implemented by addition to
Eq. (\ref{lmax}) of a matter Lagrangian, ${\cal L}_m$, derived from the free matter
Lagrangian by the minimal substitution, $p_\mu \rightarrow p_\mu - eA_\mu$.
Clearly, ${\cal L}_m$ also is not gauge-invariant.
Although corrective
procedures\cite{bel40} can be found in any textbook on field theory,
it would be preferable to avoid these problems from the beginning. 

In this paper the Fermi Lagrangian is explored,
\begin{equation}
{\cal L}_F = { 1 \over 2} \epsilon_0 c^2 \partial_\mu A_\nu \partial^\mu A^\nu 
\label{lfer}
\end{equation}
as an alternative to Eq. (\ref{lmax}). 
This Lagrangian gives rise to wave equations for $A_\mu$, which
were employed in Fermi's pioneering paper on field quantisation\cite{fermi}.
${\cal L}_F$ and its associated Noether currents are occasionally
discussed\cite{achieser,jauch,bogol,mandl,cohen} in the context of covariant
quantisation, but to my knowledge no study exists
of the full consequence of ${\cal L}_F$
as the starting point of electromagnetic field theory.

It will first be shown how the wave equation, 
furnished with boundary conditions at infinity,
uniquely determines the field
and that charge conservation leads to the Lorenz condition.
Then the conservation laws are discussed and a spin
operator is defined. Subsequently, minimal coupling is derived from charge
conservation and the force density is discussed. A simple application of the
formalism is given. Finally, covariant quantisation is discussed and it is
argued that there is no need for a redefinition of the metric.

Consider a Lagrangian consisting of a field and a matter part
with minimal substitution, that is ${\cal L}= {\cal L}_F + {\cal L}_m$.
The field equations of motion are the inhomogeneous wave equations 
\begin{equation}
\partial_\lambda \partial^\lambda A_\mu = - \mu_0 j_\mu
\label{wave}
\end{equation}
with
\begin{equation}
- \mu_0 j_\mu = \partial {\cal L}_m / \partial A^\mu .
\label{current}
\end{equation}
Now adopt as a boundary condition that
$A_\mu=0$ in a point P=(r,t) if $j_\mu=0$ in the entire light cone of P.
The derivatives of $A_\mu$ also vanish,
if $j_\mu=0$ in the union of the light cones of
an infinitesimal environment of P. 
With this,
Eq. (\ref{wave}) fully determines $A_\mu$ in the point P if $j_\mu$ is specified 
in the entire light cone of P,
for the difference of two fields must vanish if they correspond
to the same current distribution in the entire light cone of P.
An equivalent statement of the boundary condition is
that a source at P does not contribute to the potential outside the lightcone of P.

Maxwell's equations follow from Eq. (\ref{wave}) and the Lorenz condition,
$\partial_\mu A^\mu=0$.
It is now shown that the Lorenz condition follows from charge conservation,
$\partial_\mu j^\mu=0$,
which in turn requires that ${\cal L}_m$ is real. 
Indeed, from this and Eq. (\ref{wave}) it follows that 
$\partial_\lambda \partial^\lambda {\partial_\mu A^\mu}=0$
in the entire Minkowski space, or,
in k-space, $k_\lambda k^\lambda k_\mu A^\mu (k)=0$.
This implies $k_\mu A^\mu=0$
or $k_\mu k^\mu=0$, or both. 
If $k_\mu k^\mu=0$ then $A_\mu$ is a solution 
of the free wave equation in entire Minkowski space
and must vanish due to the boundary condition.
This leaves the possibility that $k_\mu A^\mu (k)=0$, 
so that $\partial_\mu A^\mu=0$. 
Thus the Lorenz condition is fulfilled if and only if charge is conserved.
It is a consequence of a property of matter and in no way an intrinsic
restriction of the degrees of freedom of $A_\mu$.
This also means that,
whereas in the standard (gauge invariant) theory 
"non-conservation of charge is inconceivable" \cite{schwinger},
this is not the case here.

Another consequence of charge conservation is minimal coupling.
If the charged matter is described by a complex field $\psi$,
the conserved current is given by
\begin{equation}
j_\mu = {e \over 2i} {{\partial {\cal L}_m} \over {\partial ( \partial_\mu \psi ) }} \psi + c.c.
\end{equation}
This expression is equal to Eq. (\ref{current}) only if
$\partial_\mu \psi$, $\partial_\mu\psi^*$ and $eA_\mu$ occur in ${\cal L}_m$ in the
combinations $i \partial_\mu \psi - eA_\mu$ and $i\partial_\mu \psi^* + eA_\mu$.
Thus minimal coupling is necessary and sufficient to assure
current conservation in Eq. (\ref{current}).

The field conservation laws are obtained in a straightforward manner
by application of Noether's theorem to ${\cal L}_F$.
The field energy-momentum density is 
\begin{equation}
T_f^{\mu\nu} =
\epsilon_0 c^2  \partial^\mu A_\rho \partial^\nu A^\rho - g^{\mu\nu} {\cal L}_F
\label{emdens}
\end{equation} 
$T_f$ is symmetric under exchange of its indices,
as required by the proportionality
of energy flux and momentum density (Ref. \cite{landau}, \S 32). 
Eq. (\ref{emdens}) also coincides 
with the definition of the energy-momentum tensor in general
relativity, $ \partial {\cal L} / \partial g_{\mu\nu} $,
which is always symmetric. ${\cal L}_F$ is the unique Lagrangian that
yields an energy momentum tensor with these properties. It is important to note
that the total energy-momentum, $\int d^3x (T_f^{0\mu} + T_m^{0\mu})$, 
including the matter
contribution, has the same value as in the standard theory. 
Still, the  distribution of energy-momentum, 
which is not an experimentally accessible quantity, is different.
Moreover a quantity of energy-momentum of $\int d^3x j^0 A^\mu$ that is
attributed to the {\it matter} energy-momentum in the standard theory
is attributed to the {\it field} energy-momentum here. As a consequence
the energy density of the field is not positive definite. 
Notably, an electrostatic field exhibits a negative energy density.
The momentum density of static fields strictly vanishes,
whereas the Poynting vector ascribes momentum density 
to crossed static electric and magnetic fields\cite{panofsky}.

The different partition of energy-momentum density between matter and field
yields a density of force exerted on the matter that differs from the Lorentz force, namely
\begin{eqnarray}
f_m^\mu &=& \partial_\lambda T_m^{\lambda\mu} \nonumber \\
&=& - \partial_\lambda T_f^{\lambda\mu} \nonumber \\
&=& - j_\lambda \partial^\mu A^\lambda.
\label{force}
\end{eqnarray} 
As an example, consider a system of interacting classical point charges.
The Lorentz force on a single particle equals
the rate of change of $d(p_i^\mu)/ds = m d(u_i^\mu)/ds$.
However, in the presence of a field 
this does not correspond to a conserved momentum
so that the Lorentz force does not obey Newton's third law\cite{newton}.
On the other hand, the force given by Eq. (\ref{force}) equals
the rate of change of the conserved\cite{radiation} total particle momentum 
$\sum_i ( mu_i^\mu - qA_\mu(r_i) )$
and does obey Newton's third law.
The Lorentz force law follows from Eq. (\ref{force})
by setting the rate of change of particle momentum,
$d(mu_i^\mu - qA^\mu(r_i))/ds = j_\lambda \partial_\mu A^\lambda (r_i)$, 
equal to the r.h.s. of Eq. (\ref{force}). 

The field angular momentum density is
\begin{equation}
J_f^{\lambda\mu\nu} = x^\mu T_f^{\lambda\nu} - x^\nu T_f^{\lambda\mu} 
+ \partial_\lambda A^\mu A^\nu - \partial_\lambda A^\nu A^\mu .
\label{ang}
\end{equation}
The first two terms on the r.h.s. represent the field orbital
momentum density, which is conserved because of the symmetry of $T_f^{\mu\nu}$.
The remaining terms represent spin angular momentum,
which is separately conserved because
${\cal L}_f$ is invariant under Lorentz transformations of $A_\mu$ alone.
In fact, this interpretation fulfills the requirement \cite{enk}
that the total angular momentum reduces to spin in the rest frame.
The rest frame of a field can be defined
as the one in which the total momentum vanishes.
For a massless travelling wave the momentum can be made arbitrarily small
by a suitable choice of frame, but not vanish.
The orbital angular momentum vanishes quadratically,
while the spin vanishes linearly.
If the field is normalised to have energy $\hbar\omega$,
the spin equals $\pm \hbar$ and the orbital angular momentum vanishes linearly.
For a standing wave the interpretation is completely straightforward: 
both linear and orbital angular momentum vanish
so what remains can only be spin.

\begin{figure}
\vskip 0.5cm
\epsfbox{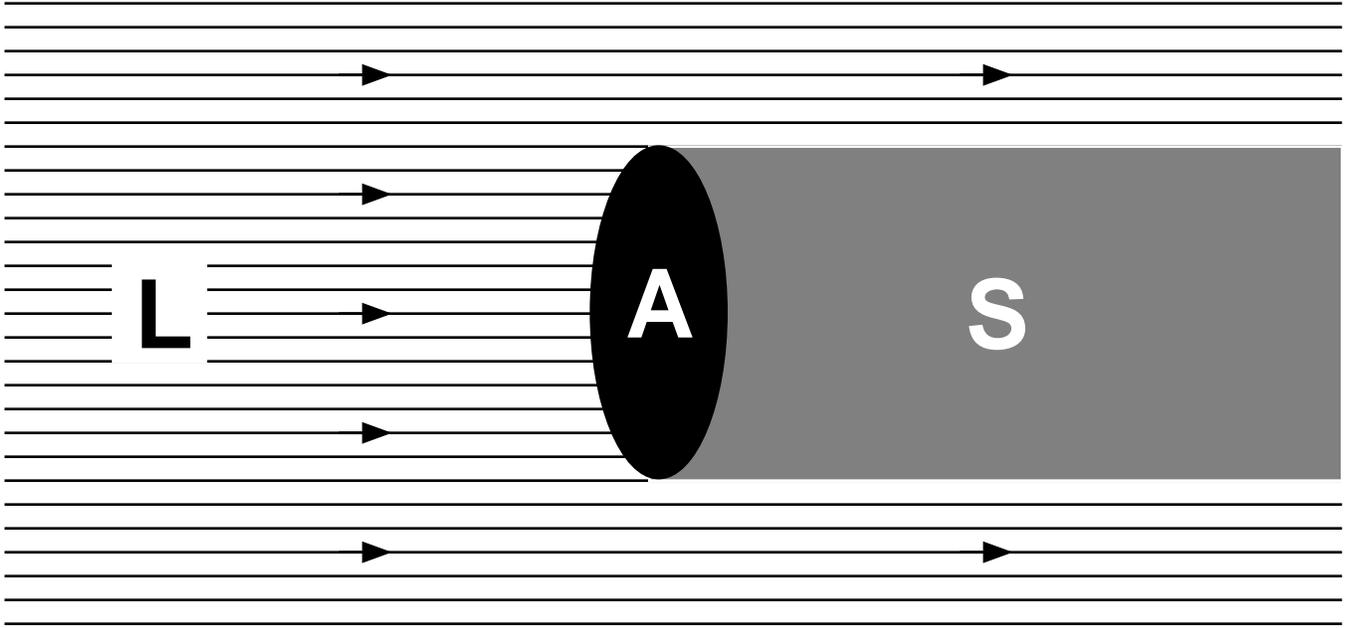}
\vskip 0.5cm
\caption{When an absorbent disk (A) is placed in a plane 
circularly polarised wave (L) casting a shadow (S),
it will experience a torque.
In the present approach this equals the rate at which spin density is absorbed.
In the standard approach angular momentum is conserved
because a sharply localised negative spin density forms
at the edge of the shadow region.}
\vskip 0.5cm
\label{fig:spin}
\end{figure}

As an example consider a plane wave with circular polarisation,
\begin{eqnarray}
A^x&=&Acos(\omega t-kz) \nonumber \\
A^y&=&Asin(\omega t-kz) .
\label{plane}
\end{eqnarray}
The angular momentum density is equal to $S^z=S^{xy} = \omega A^2$.
In the standard theory, on the other hand,
only orbital angular momentum density exists
and, since the Poynting vector is along $\vec{z}$,
its z-component vanishes. 
The total spin is equal and the difference is recovered
in regions of space where the wave is not plane
and consequently the Poynting vector no longer parallel to the z-direction.
If a light absorbing disc with cross
section D is placed in the bundle, as shown in Fig. \ref{fig:spin},
it will experience a torque $N^z$ \cite{beth}.
$N^z$ is equal to the rate of spin absorption, so that $N^z = c S^z D$.
The same reasoning would yield {\it zero} torque in the standard theory, 
because the z-component of angular momentum density vanishes.
In the standard approach the torque is associated
with field angular momentum generated {\it behind} the absorber.
At the edge of the cylindrical shadow cast by the disk
a sharply peaked distribution of angular momentum pointing in the negative
z-direction occurs. Clearly, the standard description is far from straightforward.

It is now straightforward to define spin operators.
The expression for the total spin, 
\begin{eqnarray}
S^{\mu\nu} &=& \int d^3x S_0^{\mu\nu} \nonumber \\
&=& 
\int d^3x (\partial_0 A^\mu) A^\nu - (\partial_0 A^\nu) A^\mu ,
\label{spin}
\end{eqnarray}
has the form of an expectation value of the operator
\begin{equation}
(g^\nu_\alpha g^\mu_\beta - g^\nu_\beta g^\mu_\alpha) \partial_0 .
\end{equation}
Clearly, eigenfuctions of this operator should be stationary,
that is they should be eigenfunctions of $\partial_0$.
For a real vector field this condition
has to be relaxed in the following way:
1) $A^\mu$ is an eigenfunction of $\partial_0^2$ and
2) $\partial_0$ shifts the phase of $A^\mu$ by ${\pi \over 4}$
in addition to multiplication by $\omega$.
Denoting the phase shift operation by $R$ one has $\partial_0 A^\mu = \omega R A^\mu$.
The factor $\omega$ can be absorbed into the normalisation of the field. 
Notably, if the field is normalised to carry energy $\hbar\omega$ 
the factor $\omega$ combines with the normalisation
$\sqrt {\hbar \over \omega}$
to give a factor $\hbar$.
For simplicity $\hbar$ is now set to unity.
The operator $R$ replaces the imaginary number $i$,
which is the ${\pi \over 4}$ operator in the Klein-Gordon case.
For example, one has $R^2 = -1$.
For a particular propagation direction the spin operator can now be written as
\begin{equation}
S^{\mu\nu} = { 1 \over R } ( g^\nu_\alpha g^\mu_\beta - g^\nu_\beta g^\mu_\alpha ).
\label{spinop}
\end{equation}
The commutation relations are
\begin{eqnarray}
[S^{\mu\nu},S^{\rho\sigma}] =
{ 1 \over R} 
&( g^{\nu\rho}S^{\mu\sigma} + g^{\mu\sigma}S^{\nu\rho} \nonumber \\
&- g^{\mu\rho}S^{\nu\sigma} - g^{\nu\sigma}S^{\mu\rho} ) .
\label{comrel}
\end{eqnarray}
These contain the proper commutation relations
of the 3D pseudovector operator $S_i = \epsilon_{ijk} S_{jk}$.
One can define $S^+$ and $S^-$ operators with $R$ replacing $i$.
For the circularly polarised wave $|A>$ propagating along $\vec z$
given by Eqs. (\ref{plane}) 
one has $S^z |A> = |A>$.
It is straightforward to check that $S^+|A> = 0$
and that $(S^-)^2|A>$ is an eigenfunction of $S^z$ with $-1$.
One has $S^-|A> = \sqrt{2} (0,0,0,sin(\omega t - kz))$,
which is an eigenfunction of $S^z$ with vanishing eigenvalue.
This longitudinal wave belongs to the Hilbert space
but cannot be observed because of charge conservation.
It is easily verified that similar conclusions hold 
for the general eigenfunction of $S^z$,
$A^\mu= \hat{x}(
sin (\omega t) f( \vec{k} \cdot \vec{r} ) + 
cos (\omega t) g( \vec{k} \cdot \vec{r} )) +
\hat{y}(cos (\omega t) f( \vec{k} \cdot \vec{r} ) -
sin (\omega t) g( \vec{k} \cdot \vec{r} )) $.
Eqs. (\ref{spinop}) and (\ref{comrel}) define a second set of spin operators,
$\tilde{S}_i = S_{0i}$,
with eigenfunctions that cannot be observed because of charge conservation. 

Now that the validity of the classical theory has been established, 
the problems encountered in the covariant field quantisation procedure\cite{cohen},
are discussed.
Quantisation is achieved by interpreting the total free field energy,
\begin{eqnarray}
H_F &=&
\int d^3x T_f^{00} \nonumber \\
&=&  { 1 \over 2} \epsilon_0 c^2 \sum_{i=0}^z \int d^3x
\left( ( \partial_i \vec{A} )^2 - ( \partial_i A^0 )^2 \right)
\end{eqnarray}
as the hamiltonian of a collection of harmonic oscillators.
The commutation relations between the creation and annihilation operators are
\begin{equation}
[a_\mu (m),a^+_\nu (m')] = g_{\mu\nu} \delta (m-m'),
\label{com} 
\end{equation}
where $m$ is a discrete or continuous variable designating the field mode created by $a(m)$.
For a travelling plane waves $m$ stands for $\vec{k}$.
Eq. (\ref{com}) implies that 
\begin{equation}
[a_0(m),a^+_0 (m')] = - \delta (m-m'),
\label{com00}
\end{equation}
which gives 
\begin{equation} 
<a^+_0 (m) \psi_0 | a^+_0 (m') \psi_0 > ~ \leq 0,
\end{equation}
where $\psi_0$ is the ground state.
This constitutes a contradiction if Eq. (\ref{com}) is interpreted
as a relation between {\em components} of $A^\mu$ and $g^{\mu\nu}$.
The problem disappears, however, if $a^{+\mu} (k)$ and $a^\mu (k)$
are interpreted as Lorentz vectors with $\mu$-polarisation.
In this case Eqs. (\ref{com}) and (\ref{com00}) involve a scalar product.
The distinction between a vector and its components
can be emphasized by denoting
$a^{+(\mu)} (m)$ and $a^{(\mu)} (m)$
This suggests the possibility to define scalar operators $a(m)$ by
\begin{equation}
a^{(\mu)} (m) = e^{(\mu)} a(m)
\label{adef}
\end{equation}
and likewise for $a^+(m)$.
Here the $e^{(\mu)}$ are unit vectors that are obtained from $g^{\mu\nu}$
by fixing one of the indices and obey
\begin{equation}
e^{(\mu)}_\alpha  e^{(\nu)\alpha}  = g^{\mu\nu} .
\label{unit}
\end{equation}
From the postulate
\begin{equation}
[a(m),a^+(m')] = \delta(m-m'),
\label{com2}
\end{equation}
combined with definition (\ref{adef}) and Eq. (\ref{unit}), Eq. (\ref{com}) is recovered.
This procedure has the advantage that
the commutativity of orthogonally polarised fields
follows from Eq. (\ref{unit}) and needs not to be postulated.
The norm of a photon state is positive definite, since from
\begin{eqnarray}
<a^{+(0)} (m) \psi_0 | a^{+(0)} (m') \psi_0> &=&  \nonumber \\
g^{00} <a^+ (m) \psi_0 | a^+ (m') \psi_0> &\leq& 0 ,
\end{eqnarray}
it follows that 
\begin{equation}
<a^+ (m) \psi_0 | a^+ (m') \psi_0>  \quad \geq 0 .
\end{equation}
It is easily checked that the photon number operator is
\begin{equation}
N (m) = a^+(m) a(m) .
\end{equation}
The probability interpretation of quantum mechanics is thus maintained,
as well as Lorentz covariance,
and the notorious 'indefinite metric' problem is avoided.
The Lorenz condition is implemented 
simply by using only polarisations that obey
\begin{equation}
k_\mu \epsilon^\mu = 0,
\label{lorenz}
\end{equation}
where $\epsilon^\mu$ is the polarisation of the potential.

In summary, it was shown that the Lagrangian
${\cal L}_F = { 1 \over2} \epsilon_0 c^2 \partial_\mu A_\nu\partial_\mu A_\nu$ 
leads to a valid theory of electromagnetism
with a uniquely defined potential.
The Lorenz condition and minimal coupling emerge 
as consequences of charge conservation.
The canonical spin density is covariant and conserved
and permits the definition of a photon spin operator.
The 'indefinite metric' problem
encountered in the covariant quantisation procedure is avoided.


\begin{references}
\bibitem{radiation}
The absence of radiation effects is assumed.
\end{references}
\end{document}